\documentclass[12pt]{article}
\usepackage{mathtext}
\usepackage{graphicx}
\usepackage[T2A]{fontenc}
\usepackage[utf8]{inputenc}
\usepackage{amsfonts}
\usepackage{amssymb}
\usepackage{amsmath}
\usepackage{pgfplots}
\usepackage{braket}
\usepackage{tikz}
\usepackage{float}
\usepackage{multicol}
\newcommand{\g}{\gamma}

\newcommand{\w}{\omega}
\newcommand{\s}{\sigma}

\renewcommand{\a}{\alpha}

\begin{document}
\title{About chemical modifications of finite dimensional models of QED}

\author{Vitaliy Afanasyev$^{1}$, Zheng Keli$^{1}$, Alexei Kulagin$^{1}$, Hui-hui Miao$^{1}$,\\  
Yuri Ozhigov$^{1,2}$, Wanshun Lee$^{1}$, Nadezda Victorova$^{3}$\\
{\it 
1.Moscow State University of M.V.Lomonosov, VMK Faculty, Russia
} \\
{\it 2. Institute of Physics and Technology RAS (FTIAN), Moscow}\\ 
{\it 3. Russian State University for the Humanities
}}

\maketitle

\begin{abstract}
    Suggestion of modifications of finite-dimensional QED models are proposed for interpreting chemical reactions in terms of artificial atoms and molecules on quantum dots placed in optical cavities. Moving both photons and atoms is possible between the cavities. Super dark states of diatomic systems are described, in which the motion of atoms between cavities is impossible due quantum interference. Chemical processes with two level atoms and three level atoms with lambda spectrum are schematically modeled by solving the single quantum master equation with the Lindblad operators of photon leakage from the cavity and influx into it; association and dissociation reactions then differ only in the initial states. An example is given of the optical interpretation of the transition of an electron from atom to atom in terms of the multilevel Tavis-Cummings-Hubbard model with an estimate of the accuracy. Polyatomic chemical reactions are too complex for accurate modeling. Our method of rough interpretation helps to obtain their long-term results, for example, the form of stationary states of reagents, such as dark and super dark states.

\end{abstract}

\newpage
\tableofcontents
\newpage

\section{Introduction}
    One of the main tasks of mathematical modeling of natural phenomena - the predictive modeling of chemistry attracts increasing interest due to the growing capabilities of supercomputers. They allow simulation of limited molecular structures in the framework of ''quantum chemistry'' - stationary states of molecules. From recent publications, we will mention the work \cite{1}, where authors propose the package written on $C^{++}$ which advantage is the visual representation of molecules and quick finding their energy curve. Quantum method open new perspectives in simulating chemistry even before the building of the full scale quantum computer (\cite{2}). 

However, there is still no computer simulator of the dynamics of chemical reactions. The stumbling block here is the fundamental role of the electromagnetic field, whose representation requires the lion's share of computing resources, which already grow exponentially with the increase in the number of particles.

 This presupposes a quantum description of reactions with the interaction of the charges of electrons and nuclei with the field. The traditional scheme of mathematical modeling ''computer - real system'' is not suitable for such a task, as the complexity here grows as an exponential of the number of real particles, and in the presence of a field, the complexity becomes even greater.

\begin{figure}
        \centering
        \includegraphics[scale=0.7]{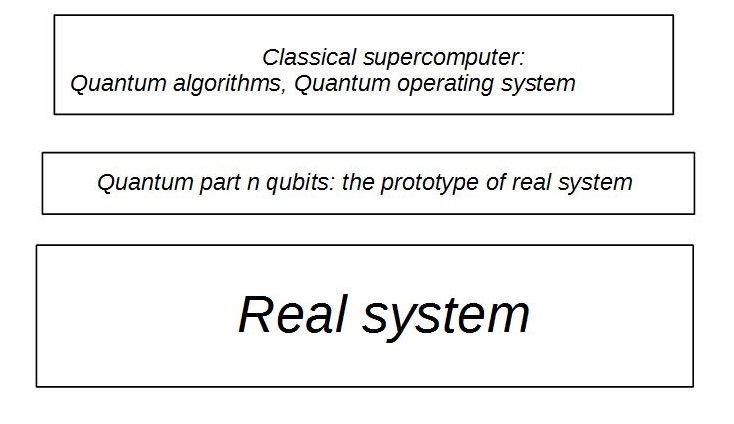}
        \caption{Simulation of reality at the quantum level: general scheme}
        \label{fig:1}
     \end{figure}

    Therefore, it is necessary to add an intermediate element to the computer simulation scheme - the quantum part. This is an array of qubits organized in some way, which imitates the behavior of the considered system of atoms and the processes of their recombination into molecules. A classical supercomputer is loaded with a quantum operating system that controls the quantum part of the computer; we learn about the result of the simulation by measuring its state. This is a quantum computer designed for the study of chemistry, its schematic is shown in Figure \ref{fig:1} (for the mode detailed scheme see, for example, \cite{3}). The main task of such a computer is to determine the result of reactions. Reproduction of accurate dynamics for complex processes is hardly possible. However, we can hope that this model, due to its simplicity, will allow us to determine the final products of the reaction and the influence of certain external factors on them, such as entangled states of photons or nuclear spins.
    
    Since the main difficulty represents the description of field, for the ''dynamical chemistry'' we have to use the models similar to  cavity QED finite dimensional models. In the last time a lot of research has been carried out in the area of modifications of Jaynes-Cummings-Hubbard (JCH) and Tavis-Cummings-Hubbard model (TCH): studying phase transitions \cite{JCH-phase-diagram}\cite{JCH-three-body-interactions}, search for metamaterials \cite{JCH-supersolid}, studying quantum many-body phenomena \cite{JCH-many-body}, etc.

We will use the ''chemical'' modification of TCH model, directed at the quantum part of computer as optical cavities, connected with optical waveguides and the special bridges for the transition of atoms betwee cavities. This model is schematic, and we do not consider the technical problems related to its construction; though the working optico-mechanical quantum devices preserving the coherence for the moving atoms show that this scheme is realistic.

\section{Finite dimensional models of electrodynamics: dark states}

Experiments on individual atoms and ions dictate the use of optical cavities and their varieties
(\cite{3}, \cite{4},\cite{5}). Describing quantum part of ''chemical computer'' we use the traditional cavity QED models, modified for our aims. 

    The main mathematical model of cavity QED is TCH model, which describes the dynamics of two level atoms and one mode field in the optical cavities connected with the optical fiber. Each cavity containes a few identical atoms and holds photons of the mode corresponding to the atomic excitation. Photons of this mode in each fixed cavity are indistinguishable, they interact with atoms inside of the cavities and can jump from one atom to the other along the waveguides. The Hamiltonian of TC model for one cavity has the form:
    
    \begin{equation}
        H_{TC}=H_c+H_a+H_i,\ H_c=\hbar\omega a^+a,\ H_a=\hbar\omega\sum\limits_{j=1}^n\s_j^+\s_j,\ \ H_i= (a^++a)(\bar\s^++\bar\s),
        \label{TC}
    \end{equation}

    where upper index $+$ means comjugation, $a^+,a$ are the field operators of creation and annihilation of a photon, $\bar\s=\sum\limits_jg_j\s_j$, $\s^+_j,\s_j$ - operators of excitation and relaxation of $j$th atom, and $g_j$ is its energy of interaction with the field.
    For the weak interaction $ g_j/\hbar\w\ll 1$ we can use RWA - rotating wave approximation, and simplify the interaction term $H_i$ replacing it with $H_i^{RWA}=(a\bar\s^++a^+\bar\s)$. 
    
    Hamiltonian of TCH model for teh set of cavities in the exact form (or in RWA) has the form:

    \begin{equation}
        \label{TCH}
        H_{TCH}=\sum\limits_{i=1}^mH^i_{TC}+\sum\limits_{1\leq i<j\leq m}\mu_{ij}(a_i^+a_j+a_ia_j^+).
    \end{equation}

    The complete description of eigenstates of TC model is complex (see \cite{Tav}), however, there is practically significant type of eigenstates, which atomic part is dark states of atoms. In such a state atoms cannot emit a photon though its energy is nonzero. 
    In RWA dark subspace is $Ker(\bar\sigma)$, $\bar\sigma=\sum\limits_ig_i\sigma_i$, in the exact TCH model dark subspace is $D_n=Ker(\bar\sigma+\bar\sigma^+)$.
    
    Dark subspace in the exact model exists only is all $g_i=g$ and $n$ is even, then $dim(D_n)=C^{n/2}_{n}-C^{n/2+1}_{n}$. 
    Any dark state in the exact model is a linear combination of tensor products of singlets $\frac{1}{\sqrt 2}(|01\rangle-|10\rangle).$ (see \cite{Oz}).
    \bigskip
    
    For ensembles of $d$-level atoms the analogous statement is numerically verified only for $d=3,\ n\leq 20$. Here singlets are replaced my multi-singlets: \\
    $\frac{1}{\sqrt d!}\sum\limits_{\pi\in S_d}(-1)^{\sigma(\pi)}|\pi(1)\pi(2)...\pi(d)\rangle$.
    \bigskip

    We include the position of every atom:
    \begin{equation}
    \begin{array}{ll}
    |n_1,n_2,...,n_k\rangle_{ph}&|at_1state,\ at_1position\rangle|at_2state,\ at_2position\rangle\\
    &...|at_nstate,\ at_nposition\rangle,
    \end{array}
    \label{atom_jump_basic}
    \end{equation}
    
    We add to $H_{TC}^{RWA}$ the summand, which expresses  jumps of atoms from one cavity to the other

    \begin{equation}
    \label{atom_jump}
    \sum\limits_{i,1\leq j<q\leq k}r^i_{jq}(S(i)_j^+S(i)_q+S(i)^+_qS(i)_j),
    \end{equation}
    where $S_j, S_j^+$ - are the operators of annihilation and creation of $i$th atom in the cavity $j$, 

    {\bf A black state} - in which two atoms can neither move, nor emit a photon:
    \begin{equation}
    \label{chernoe}
    |C_2\rangle=|s_1\rangle-|s_2\rangle,
    \end{equation}
    where $|s_1\rangle=|01\rangle|11\rangle-|11\rangle|01\rangle$ - 2 atoms singlet in the first cavity,  $|s_2\rangle=|02\rangle|12\rangle-|12\rangle|02\rangle$ - the analogous singlet in the second cavity.

\section{Optico-mechanical model}

\begin{figure}[h]
        \centering
        \includegraphics[scale=0.8]{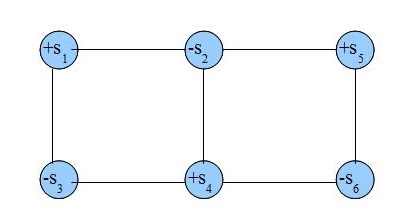}
        \caption{Even graph. Black diatomic state: $|s_1\rangle-|s_2\rangle-|s_3\rangle+|s_4\rangle+|s_5\rangle -|s_6\rangle$}.
        \label{fig:evengraph}
    \end{figure}

    An abstract model of an opto-mechanical system with moving atoms we represent by a graph of cavities connected by waveguides and transition bridges for atoms. Black diatomic states can only exist for even graphs, for which any cycle contains an even number of cavities (see Figure \ref{fig:evengraph}).
    
    \begin{figure}[h]
        \centering
        \includegraphics[scale=0.6]{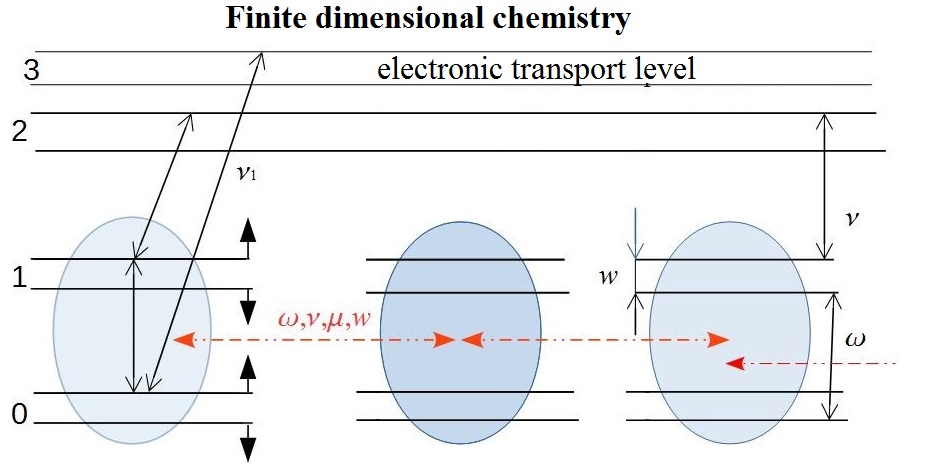}
        \caption{Artificial atoms in a cavity}
        \label{fig:chem_comp}
    \end{figure}

\section{Including of explicit electrons in separated atoms}

    This section describes our main model. It includes two or more separated atoms with two valence electrons. Each atom in this model has two level (orbits) and one transport level for electrons (see Figure \ref{fig:chem_comp}).

    According to the Pauli prohibition principle, there can be no more than one electron in each orbit. It should be noted that electrons can have two spins, so in reality the orbits should be doubled. But in this model, we assume that electrons have the same spin. Thus, there can be no more than one electron in each orbit. However, the transport level can be divided into sublevels with slightly different energies and we can assume that any number of electrons can be at the transport level. The entire system is enclosed in a multimode optical resonator that can contain all kinds of photons involved in reactions.


    We thus have 4 basic states of an atom with different electronic configurations form a basis:
    
    $$
        \left| 0 \right\rangle_{at}=\left| 0 \right\rangle_{ob_0}\left| 0 \right\rangle_{ob_1},\ 
        \left| 1 \right\rangle_{at}=\left| 0 \right\rangle_{ob_0}\left| 1 \right\rangle_{ob_1},\ 
        \left| 2 \right\rangle_{at}=\left| 1 \right\rangle_{ob_0}\left| 0 \right\rangle_{ob_1},\ 
        \left| 3 \right\rangle_{at}=\left| 1 \right\rangle_{ob_0}\left| 1 \right\rangle_{ob_1}
    $$

    Operators that describe the different types of electron transfer and interaction with the field (emission and absorption of photons):
    \begin{itemize}
        \item $T_{01at_0}=
        a_{01}^\dagger \sigma_{tp} |1\rangle\langle0|_{at_0} +
        a_{01} \sigma_{tp}^\dagger |0\rangle\langle1|_{at_0} $
        \item $T_{12at_0}=
        a_{12}^\dagger |2\rangle\langle1|_{at_0} +
        a_{12} |1\rangle\langle2|_{at_0} $
        \item $T_{23at_0}=
        a_{23}^\dagger \sigma_{tp} |3\rangle\langle2|_{at_0} +
        a_{23} \sigma_{tp}^\dagger |3\rangle\langle2|_{at_0} $
        \item (similar for atom 1) \dots
    \end{itemize}

    The Hamiltonian of this model consists of the sum of many operators, including potential energy operators, photon energy operators, and electron transfer operators, and has the form:
  $$
\begin{array}{ll}
        H =
        &P_{at_0}+P_{at_1}+P_{tp}+E_{01}+E_{12}+E_{23}
        +g_{01}T_{01at_0}+g_{12}T_{12at_0}+\\
&g_{23}T_{23at_0}
        +g_{01}T_{01at_1}+g_{12}T_{12at_1}
    +g_{23}T_{23at_1},
\end{array}
   $$
     where
    $P_{at_0}$, $P_{at_1}$ - potential energy of atoms;
    $P_{tp}$ - potential energy of the transport layer (always 0);
    $E_{01}$, $E_{12}$, $E_{23}$ - photon energy in various models;
    $T_{01at_0}$, $T_{01at_1}$, \dots - operator of excitation / relaxation of an electron upon absorption / emission of a photon.

    Such a system can have a special dark state (probably a "stable" chemical state): \\
    \begin{center}
        $
        \left| \Psi \right\rangle  = \frac{1}{\sqrt{2}}
        \left(
        \left| 1 \right\rangle_{at0}\left| 2 \right\rangle_{at1}
        - \left| 2 \right\rangle_{at0}\left| 1 \right\rangle_{at1}
        \right)
        \left| 0 \right\rangle_{tp}\left| 000 \right\rangle_{\omega_{01}\omega_{12}\omega_{23}}
        $
    \end{center}

\begin{figure}[h!]
        \centering
        \includegraphics[scale=0.4]{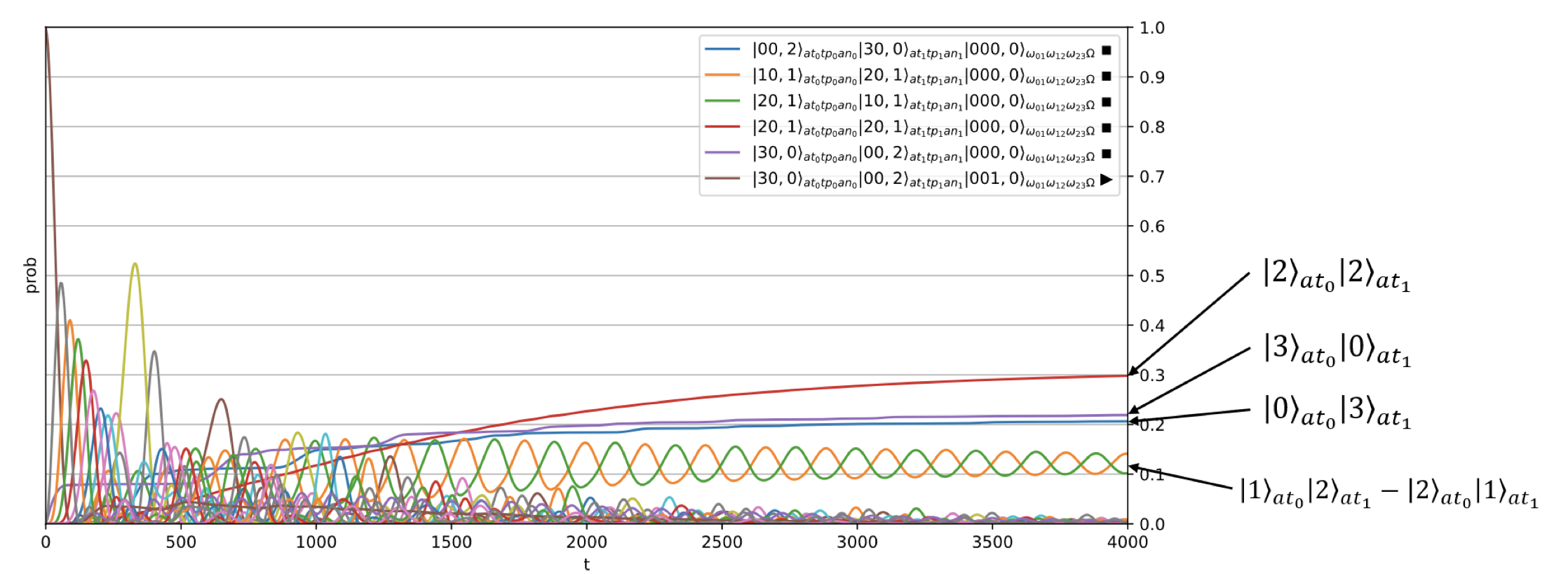}
        \caption{Optical interpretation of the charge movements}.
        \label{fig:optint}
    \end{figure}

   Thanks to the simplicity of our model we can  represent electronic jumps between atoms by only optical means (see Figure \ref{fig:optint}). Here the jump of the electron is represented by its fall to the anod level that is created especially, in three level artificial atoms, which we use instead of the chemical two level ones. The movement of electron from one cavity to the other is then roughly represented by the photon jump; appearence of the photon in the second cavity makes possible to excite the electron from anod level to the zero level. We thus have only photon transfer between atoms whereas electrons stay permanently with their atoms; only tey can fall to the lowest - anode level.

\begin{figure}[h!]
        \includegraphics[scale=0.5]{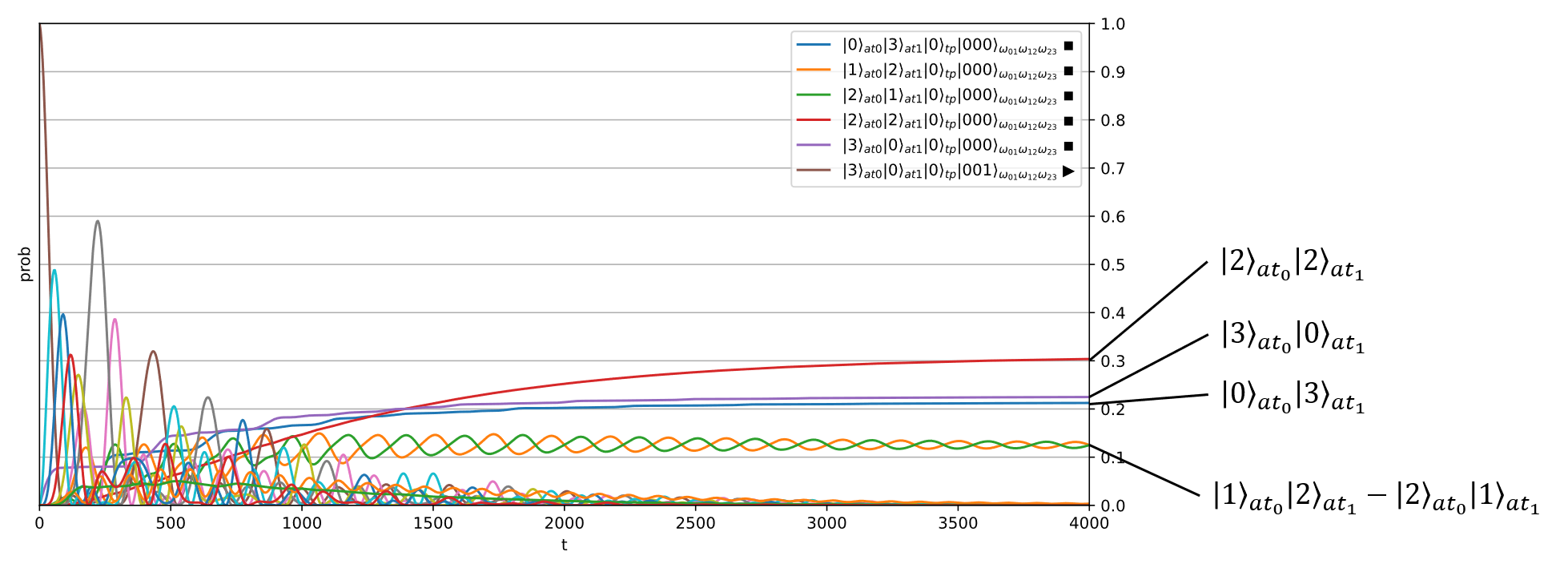}
        \caption{Evolution of the population of states over time. At the top - finite-dimensional chemistry, at the bottom - its optical interpretation
}.
        \label{fig:comparison}
    \end{figure}

Figure \ref{fig:comparison} shows the good agreement of the optical interpretation with the prototype - finite dimensional artificial chemistry. Simulation was fulfilled by quantum master equation:
$$
i\hbar\dot{\rho}=[H,\rho]+i{\cal L}(\rho),\ {\cal L}(\rho)=\sum\limits_i\g_i(A_i\rho A_i^+-\frac{1}{2}\{\rho,A^+A\})
$$ with the single type of Lindblad operators - leakage of all photons: $A_{01}\a_{01},A_{12}=a_{12},A_{23}=a_{23}$. Stabilization occures at the dark states.

\section{Hybrid spectrum in ''molecule''}
    This section describe the interaction of two atoms at different distances from each other. We are limited two extreme cases. Case A - atoms close together and case B - atoms far apart.

    \begin{figure}[h]
        \label{fig:2holes_with_spin}
        \begin{center}
            \includegraphics[scale=0.2]{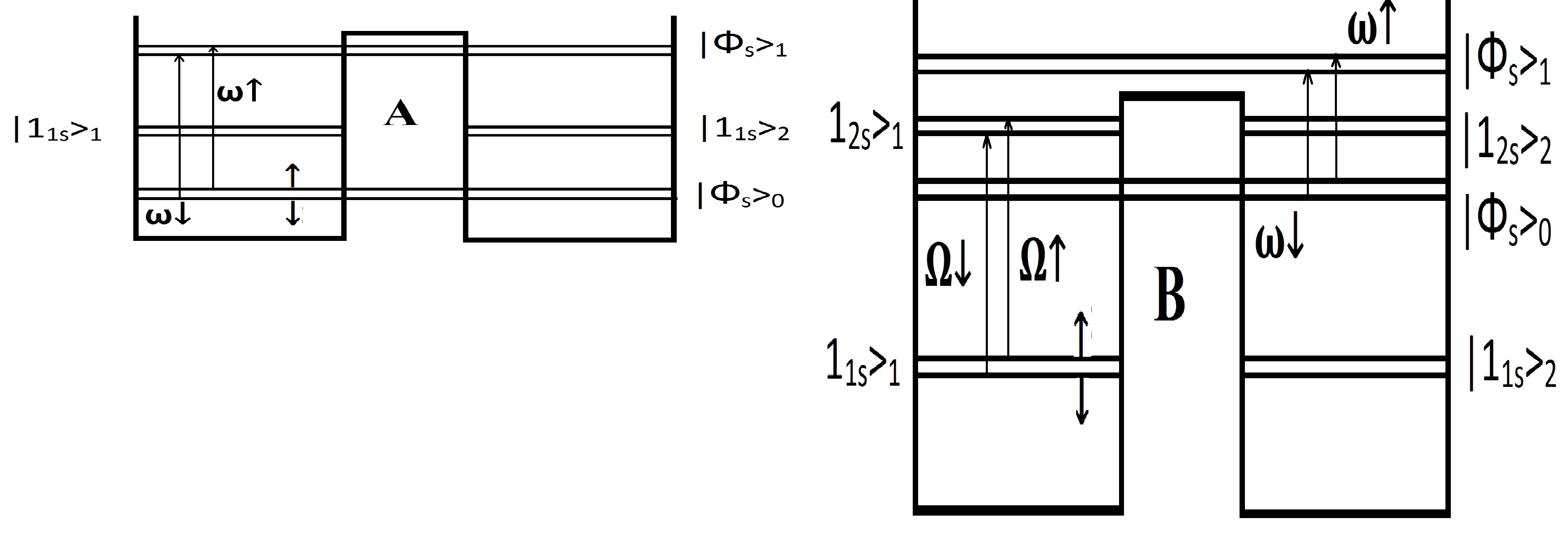}
            \caption{ {\bf A:} Two atoms at the close distance; $|\Phi_{s}\rangle_0 = \frac{|1_{1s}\rangle_1+|1_{1s}\rangle_2}{\sqrt{2}}$ $|\Phi_s\rangle_1 = \frac{|1_{1s}\rangle_1-|1_{1s}\rangle_2}{\sqrt{2}}$.
            {\bf B:} Two atoms at the far distance; $|\Phi_s\rangle_0 = \frac{|1_{2s}\rangle_1+|1_{2s}\rangle_2}{\sqrt{2}}$ $|\Phi_s\rangle_1 = \frac{|1_{2s}\rangle_1-|1_{2s}\rangle_2}{\sqrt{2}}$}
            \label{molecule}
        \end{center}
    \end{figure}
    
    Special states can be described as $ \Phi_0 $ and $ \Phi_1 $, as in the Figure ~\ref{molecule} (these are hybrid states). In this model, electrons have different spins and the photons corresponding these spins are different. 

    The model of dynamics of the single "electron" in the two hole potential induces jumps of "electron" between two holes:
    $$
    |0\rangle_{ph}\frac{|\Phi_0\rangle+|\Phi_1\rangle}{\sqrt 2}\leftrightarrow
    |0\rangle_{ph}\frac{|\Phi_0\rangle-|\Phi_1\rangle}{\sqrt 2}
    $$

    JC model describes this two hole quantum dot as the two level atom placed in the optical cavity. The standard basis of JC model: excited and basic states can be then changed to ''coordinate'' basis of holes via Hadamard transform. In the general case this is not so simple jumps of electron between two holes.

    Since the state $|0\rangle_{ph}|\Phi_0\rangle$ has zero energy, one jump from the left hole to right one means the change of the sign of the component $|0\rangle_{ph}|\Phi_1\rangle$, which evolution looks as
    $$
    |\Psi(t)\rangle = exp(-\frac{iHt}{\hbar})|0\rangle_{ph}|\Phi_1\rangle=e^{-i\w t}(cos\frac{gt}{\hbar} |0\rangle_{ph}|\Phi_1\rangle-i\ sin\frac{gt}{\hbar} |1\rangle_{ph}|\Phi_0\rangle)
    $$

    Here the coefficient $e^{i\w t }$ will play the role and the exact spatial jumps are possible if $\w\hbar/g$ has one of the forms $(1+2l)/2m$ or $2l/(1+2m)$ for integers $l,m$. The approximate jumps exists for every value of $\w\hbar/g$ by Poincare theorem; though the duration of such ''jumps'' will be very large and it looks not as Rabi oscillation.
    \bigskip

    The basis of the model can be written as follows:
    \begin{multline*}
        (|n_{\uparrow}\rangle_{\Omega})
        \otimes
        (|m_{\uparrow}\rangle_{\omega})
        \otimes
        (|n_{\downarrow}\rangle_{\Omega})
        \otimes
        (|m_{\downarrow}\rangle_{\omega})\\
        \otimes
        (|\epsilon_{1\uparrow}\rangle_1 \; , \; |\epsilon_{2\uparrow}\rangle_1)
        \otimes
        (|\epsilon_{1\uparrow}\rangle_2 \; , \; |\epsilon_{2\uparrow}\rangle_2)
        \otimes
        (|\epsilon_{1\downarrow}\rangle_1 \; , \; |\epsilon_{2\downarrow}\rangle_1)
        \otimes
        (|\epsilon_{1\downarrow}\rangle_2 \; , \; |\epsilon_{2\downarrow}\rangle_2),
    \end{multline*}
    \hfill\break
    where 
    $|n_s\rangle_{\Omega}$ - state with n $\gamma_{\Omega}$ and spin $s$, 
    $\gamma_{\Omega}$ - for moving from $|1_{1s}\rangle_a$ to $|1_{2s}\rangle_a$, 
    $|m_s\rangle_{\Omega}$ - state with m $\gamma_{\omega}$ and spin $s$, 
    $\gamma_{\omega}$ - for moving from $|\Phi_s\rangle_0$ to $|\Phi_s\rangle_1$,
    \hfill\break
    $|\epsilon_{os}\rangle_a$ - state with $\epsilon$ electrons, spin $s$, orbit $o$ and atom $a$;
    \hfill\break
    \hfill\break
    $|\Phi_s\rangle_0 = \frac{|1_{2s}\rangle_1+|1_{2s}\rangle_2}{\sqrt{2}}$ \quad $|\Phi_s\rangle_1 = \frac{|1_{2s}\rangle_1-|1_{2s}\rangle_2}{\sqrt{2}}$
    \bigskip

    The Hamiltonian of this system has a more complex structure than the Hamiltonian TC. It can be divided into four parts: two parts with total energy, part with emitting and absorption energy and part with tunneling energy:\\
    \\
    $H_{RWA} = H_{\Omega} + H_{\omega} + H_{exc} + H_{tun}$ , where
    \hfill\break
    \hfill\break
    $
    H_{\Omega} = \hbar\Omega(a^{+}_{\Omega\uparrow}a_{\Omega\uparrow} + a^{+}_{\Omega\downarrow}a_{\Omega\downarrow} + \sigma^{+}_{1\uparrow}\sigma_{1\uparrow} + \sigma^{+}_{1\downarrow}\sigma_{1\downarrow} + \sigma^{+}_{2\uparrow}\sigma_{2\uparrow} + \sigma^{+}_{2\downarrow}\sigma_{2\downarrow})
    $
    \hfill\break
    \hfill\break
    $
    H_{\omega} = \hbar\omega(a^{+}_{\omega\uparrow}a_{\omega\uparrow} + a^{+}_{\omega\downarrow}a_{\omega\downarrow} + \sigma^{+}_{tun\;\uparrow}\sigma_{tun\;\uparrow} + \sigma^{+}_{tun\;\downarrow}\sigma_{tun\;\downarrow})
    $
    \hfill\break
    \hfill\break
    $
    H_{exc} = a^{+}_{\Omega\uparrow}\overline{\sigma}_{\uparrow} + a_{\Omega\uparrow}\overline{\sigma}^{+}_{\uparrow} + a^{+}_{\Omega\downarrow}\overline{\sigma}_{\downarrow} + a_{\Omega\downarrow}\overline{\sigma}^{+}_{\downarrow}
    $
    \hfill\break
    \hfill\break
    $
    H_{tun} = g_{tun}(a^{+}_{\omega\uparrow}\sigma_{\uparrow\;tun} + a_{\omega\uparrow}\sigma^{+}_{\uparrow\;tun} + a^{+}_{\omega\downarrow}\sigma_{\downarrow\;tun} + a_{\omega\downarrow}\sigma^{+}_{\downarrow\;tun})
    $
    \hfill\break
    \hfill\break
    $
    \overline{\sigma_{s}} = g(\sigma_{s1} + \sigma_{s2}), \quad \overline{\sigma}^{+}_{s} = g(\sigma^{+}_{s1} + \sigma^{+}_{s2})
    $
    \hfill\break

    We spend two experiments for different initial states using parameters:\\
    $W = 10^{10}$; \; $w = 10^{9}$; \; $g = 10^{7}$; \; $g_{tun} = 10^{8}$; \; $\gamma_{\Omega} = 10^{8}$; \; $\gamma_{\omega} = 10^{6}$; \; $t \in [0, 10^{-5}]$
    
    \hfill\break
    {\it Experiment I.}
    \hfill\break
    Initial state: $|1_{\uparrow}\rangle_{\Omega}|1_{\uparrow}\rangle_{\omega}|1_{\downarrow}\rangle_{\Omega}|1_{\downarrow}\rangle_{\Omega} \frac{|1_{1\uparrow}\rangle_1|1_{1\downarrow}\rangle_1 - |1_{1\uparrow}\rangle_2|1_{1\downarrow}\rangle_2}{\sqrt{2}}$.\\
    Finite state:$\frac{|1_{1\uparrow}\rangle_1|1_{1\downarrow}\rangle_1 - |1_{1\uparrow}\rangle_2|1_{1\downarrow}\rangle_2}{\sqrt{2}}$

    \begin{figure}[h!]
        \begin{center}
            \includegraphics[scale=0.3]{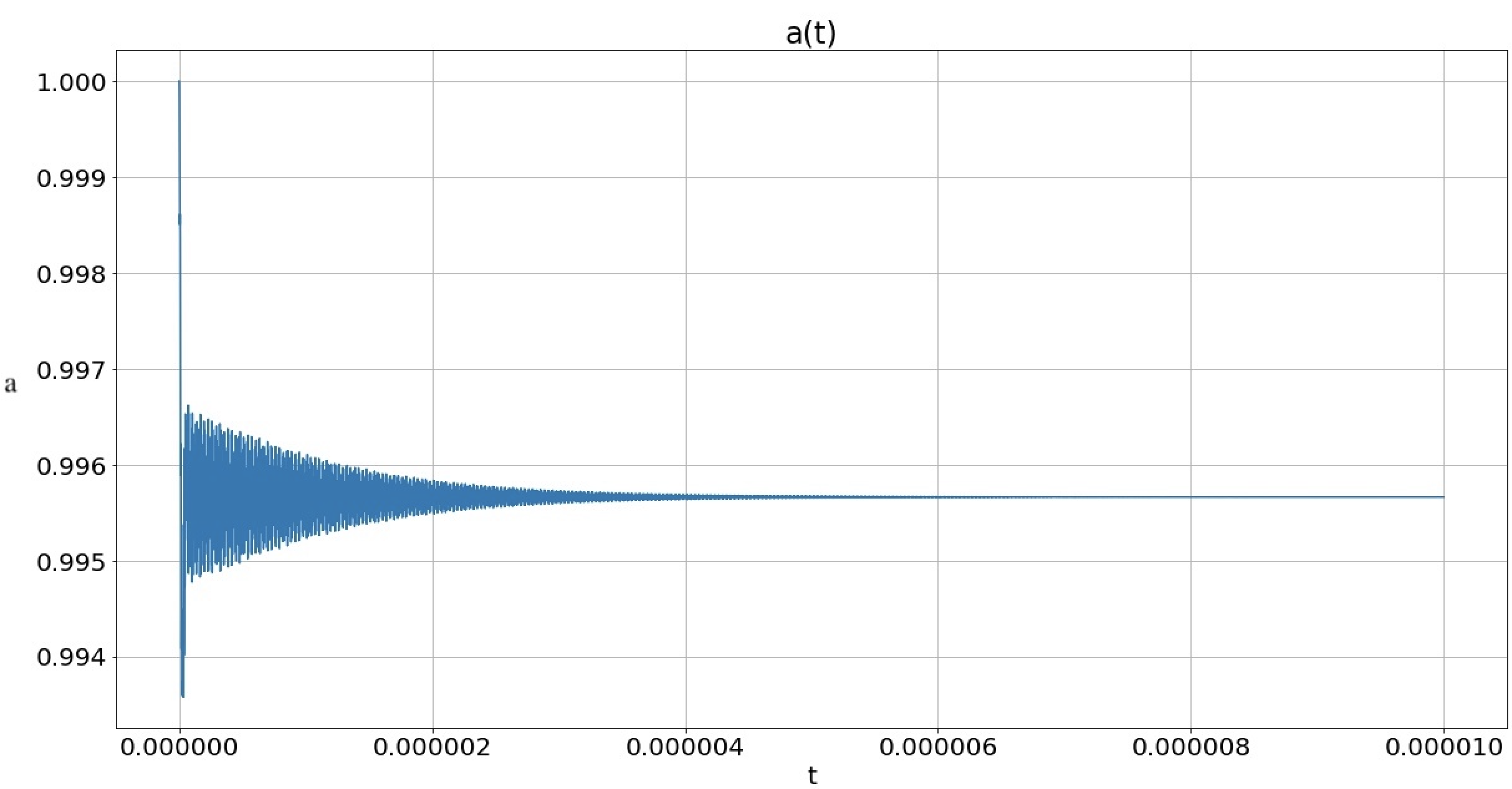}
            \caption{a(t). Experiment I. Time dependence of the degree of association.}
            \label{assosiation_1}
        \end{center}
    \end{figure}

    \hfill\break
    {\it Experiment II.}
    \hfill\break
    Initial state: $\Phi_1 = (\frac{|1_{2\uparrow}\rangle_1|1_{2\downarrow}\rangle_1 - |1_{2\uparrow}\rangle_2|1_{2\downarrow}\rangle_2}{\sqrt{2}})$\\
    Finite state: $\frac{|1_{1\uparrow}\rangle_1|1_{1\downarrow}\rangle_1 + |1_{1\uparrow}\rangle_2|1_{1\downarrow}\rangle_2}{2} +\frac{|1_{1\uparrow}\rangle_1|1_{1\downarrow}\rangle_2 + |1_{1\uparrow}\rangle_2|1_{1\downarrow}\rangle_1}{2}$

    \begin{figure}[h!]
        \begin{center}
            \includegraphics[scale=0.3]{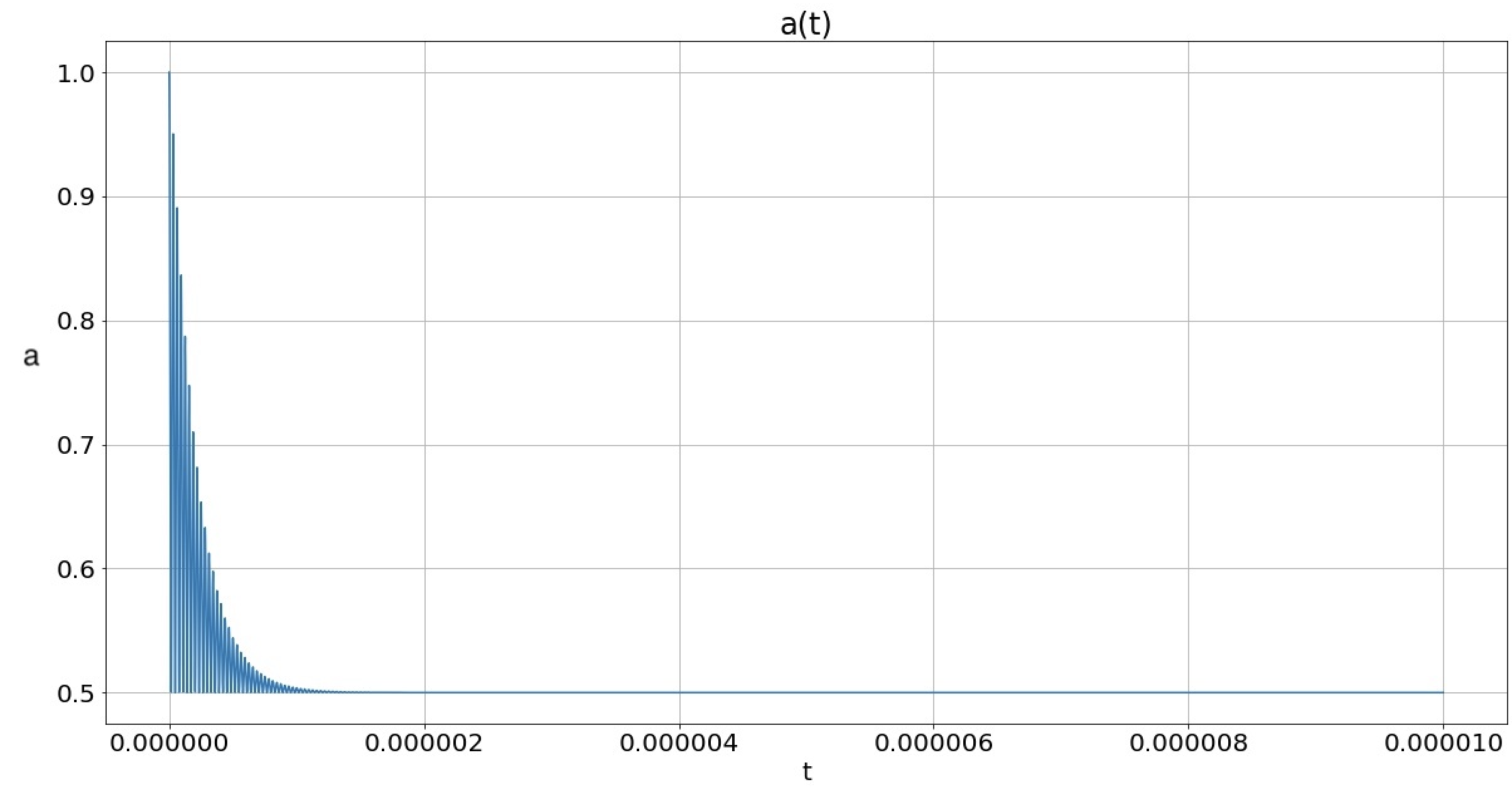}
            \caption{a(t). Experiment II. Time dependence of the degree of association. }
            \label{assosiation_2}
        \end{center}
    \end{figure}
    

    The $a$ axis represents the degree of association (Equation \ref{eqn:association}), and the $t$ axis represents the evolutionary time. You can see that in Figure ~\ref{assosiation_1} the degree of association is much greater than in Figure ~\ref{assosiation_2}. Based on the experiment carried out, it can be concluded that stable interaction between atoms is possible only in the initial dark state. The rest of the states do not lead to dark states and do not give high values of the degree of association.

\section{Simple association-dissociation model}

The detailed simulation of dissociation of the molecular ion $H_2^+$ is proposed in the work \cite{6}.  We show the radically simplified method that could be extended to much complex systems, for which the collective effects play the much more significant role than the accuracy in the single molecule description. 

    In this model we have two nuclei, two optical resonators and one electron. The excited and basic state of the electron are: $\phi_0$, $\phi_1$. Nuclei can tunnel between cavities. The association process can be described as follows: two nuclei are in different cavities, an electron absorbs a photon and become into an excited state, and then the nuclei can tunnel into one cavity, forming a molecule. Then the electron emits a photon and goes into the basic state in which the separation of atoms is forbidden, the molecule will be stable. Dissociation: In a molecule, an electron absorbs a photon and goes into an excited state, and the electron can be attracted by any proton. Therefore, the molecule becomes unstable, the nuclei tunnel into different cavities, the molecule dissociates.

\begin{figure}[h]
        \begin{center}
            \begin{minipage}[h]{0.44\linewidth}
                \includegraphics[width=1.2\textwidth]{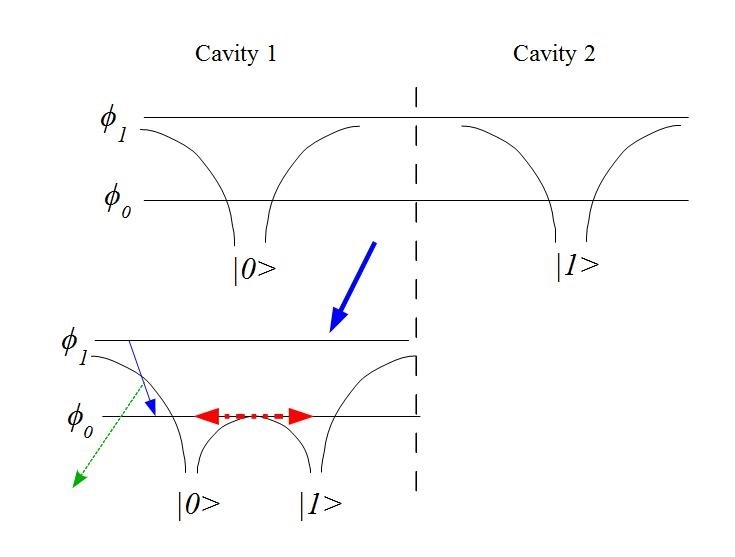} 
                \caption{Association of two artificial atoms}
                \label{fig:H12}
            \end{minipage}
            \hfill
            \begin{minipage}[h]{0.44\linewidth}
                \includegraphics[width=1.2\textwidth]{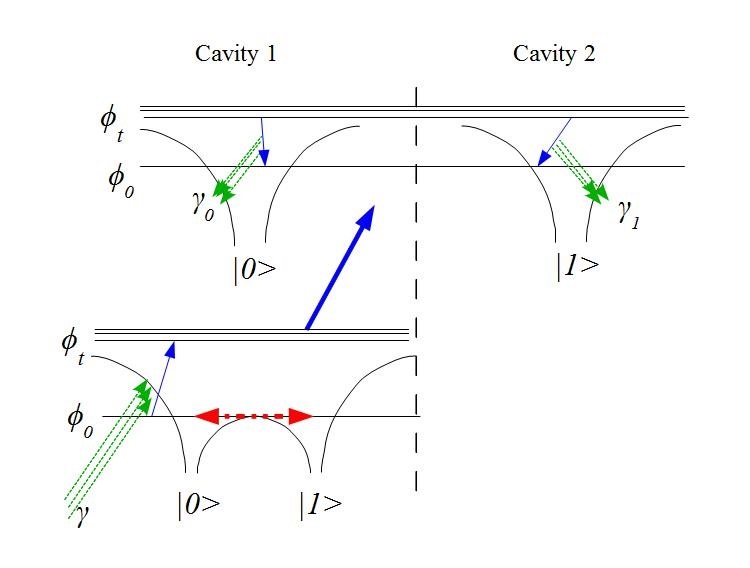}
                \caption{Dissociation of two atomic artificial molecule}
                \label{fig:H22}
            \end{minipage}
        \end{center}
    \end{figure}

    This dynamics is a simplification of real chemistry - we here ignore the reach spacial movement of nuclei, representing it by the simplest tunneling between two positions: far and close to one another. It is represented schematically at the Figures \ref{fig:H12}, \ref{fig:H22}

    Measure of association:
    \begin{equation}
        a(t) = \sum_{|\Phi\rangle}{\xi(\Phi)\langle\Phi|\rho(t)|\Phi\rangle},
        \label{eqn:association}
    \end{equation}
    it is the level of association of atoms in a state\\ with a matrix $\rho(t)$ which is a solution of the Schrödinger equation, $\xi(\Phi)$ is the level of association of basic state $\phi$,

    \begin{equation*}
        \xi(\Phi) =
            \begin{cases}
                1 &\text{if $\Phi$ is state with different charges of atoms, example $|1_{1\uparrow}\rangle_1|1_{1\downarrow}\rangle_1$}\\
                0 &\text{if $\Phi$ is state with same charges of atoms, example $|1_{1\uparrow}\rangle_1|1_{1\downarrow}\rangle_2$}
         \end{cases}
    \end{equation*}
    \hfill\break
    \hfill\break

    Basis state may be represent as: $|k\rangle_e|l\rangle_n$, where $k=0,1:|0\rangle_e$ - electron is attached by the first nuclei;\quad $|1\rangle_e$ - electron is attached by the second nuclei.\\
    $l=0,1:|0\rangle_n$ - two nucleus in one cavity;\quad $|1\rangle_n$ - two nucleus in different cavities.

    \begin{equation*}
        \phi_0=\frac{1}{\sqrt{2}}(|0\rangle_e+|1\rangle_e),\quad \phi_1=\frac{1}{\sqrt{2}}(|0\rangle_e-|1\rangle_e)
    \end{equation*}
    \hfill\break
    Hamiltonian of the system:

    \begin{equation*}
        H=H_{n.tun}+h\omega a^+a+h\sigma_e^+\sigma_e+g(a^+\sigma_e+a\sigma_e^+),
    \end{equation*}
    
    where
    \begin{equation*}
        H_{n.tun}=\sigma_e^+\sigma_e(\sigma_n+\sigma_n^+)
    \end{equation*}
    
    \bigskip

    For an association when an electron is in the $\phi_0 $ state, the photon will jump out of the cavity, so the decoherence operator:

    \begin{equation*}
        A_{e.ass}=\sigma_e\sigma_e^+a
    \end{equation*}

    For dissociation, the electron will be at two nuclei at the same time. The decoherence operator look like this:

    \begin{equation*}
        A_{e.diss1}=a\sigma_n^+\sigma_n|0\rangle_e\langle0|_e,\quad A_{e.diss2}=a\sigma_n^+\sigma_n|1\rangle_e\langle1|_e
    \end{equation*}
    
\begin{figure}[h!]
        \begin{center}
            \includegraphics[width=0.95\textwidth]{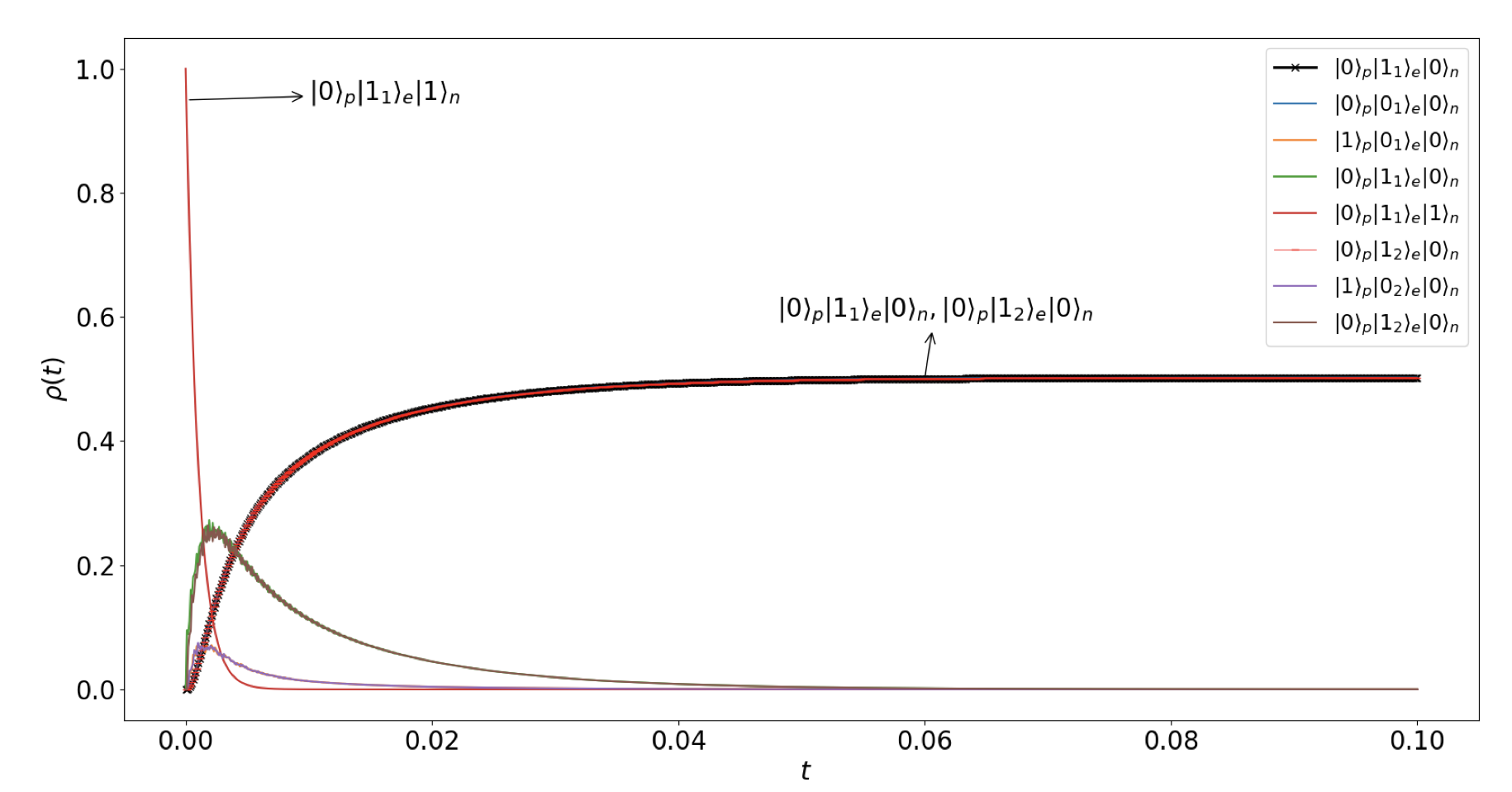}
            \caption{Association. Initial state: $|0\rangle_e|1\rangle_n$ (electron is attached by the first nuclei, two nuclei in different cavities). Finite state:$|\phi_0\rangle_e|0\rangle_n$}
            \label{association}
        \end{center}
    \end{figure}

\begin{figure}[h!]
        \begin{center}
            \includegraphics[width=0.95\textwidth]{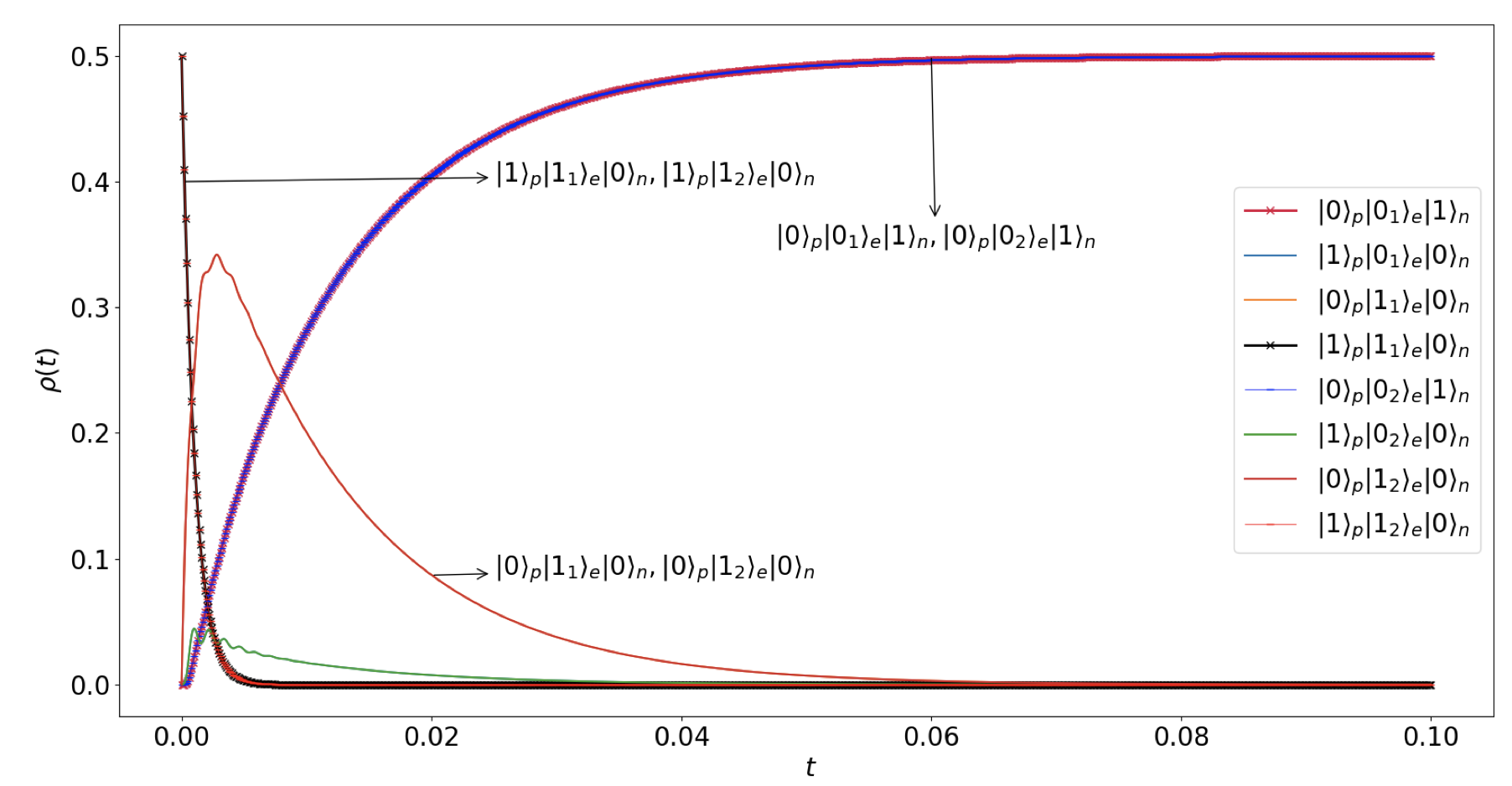}
            \caption{Dissociation. Initial state:$|\phi_1\rangle_e|0\rangle_n$. Finite state: $\frac{|0\rangle_e|1\rangle_n+|1\rangle_e|1\rangle_n}{2}$}
            \label{dissociation}
        \end{center}
    \end{figure}

    We spend two experiments for association and dissociation. They are shown on fig. \ref{association}, \ref{dissociation}.

    \newpage


\section{$\lambda$-spectrum}
	In this section, we consider a two-level atom in an optical cavity interacting with a field. This model is characterized by two decoherence factors: emission of a photon from a resonator with an intensity $\gamma_{out}$ and transformation of an atom with an intensity $\gamma_{ex}$ (possible only from an excited state $|1\rangle$)

	The physical meaning of transformation mean the decay of an atomic nucleus or the entry of an atom into a chemical reaction. In both cases, the atom loses its optical properties and cannot further interact with cavity field.

	One of our tasks is to determine the dependence of the probability of atomic transformation on the intensity of the process parameters $\gamma_{out}$ and $\gamma_{ex}$.
	
	\begin{figure}[!htbp]
        \includegraphics[width=0.58\textwidth]{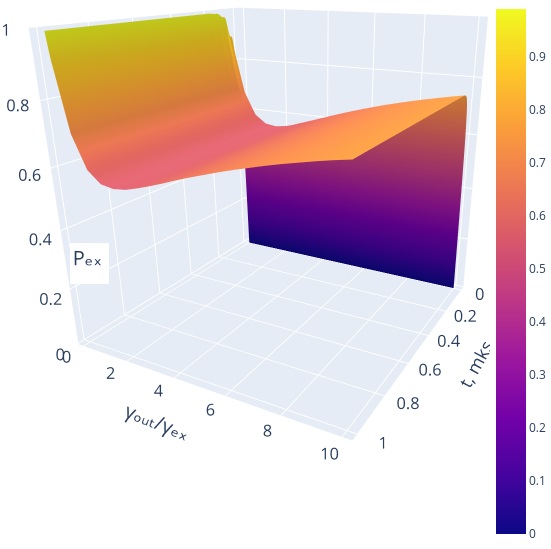}
		\includegraphics[width=0.46\textwidth]{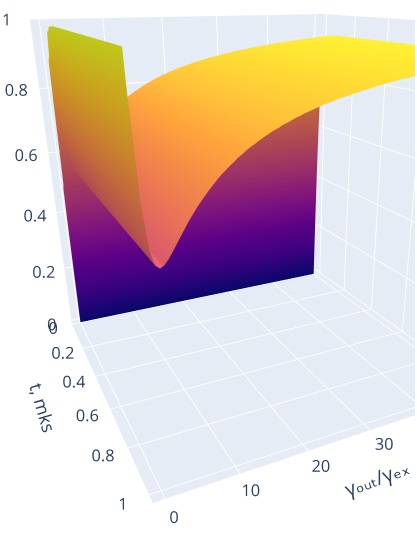}
		\caption{Probability of atomic transformation ($0 \le \gamma_{out} / \gamma_{ex} \le 10$). There is a quantum "bottleneck" effect: a counterintuitive decrease of the probability of atomic decay with the increase of its intensity $\gamma_{out}$ on some interval.}
		\label{img1}
	\end{figure}
	
	\newpage

\begin{figure}[h!]
        \begin{center}
            \includegraphics[width=0.9\textwidth]{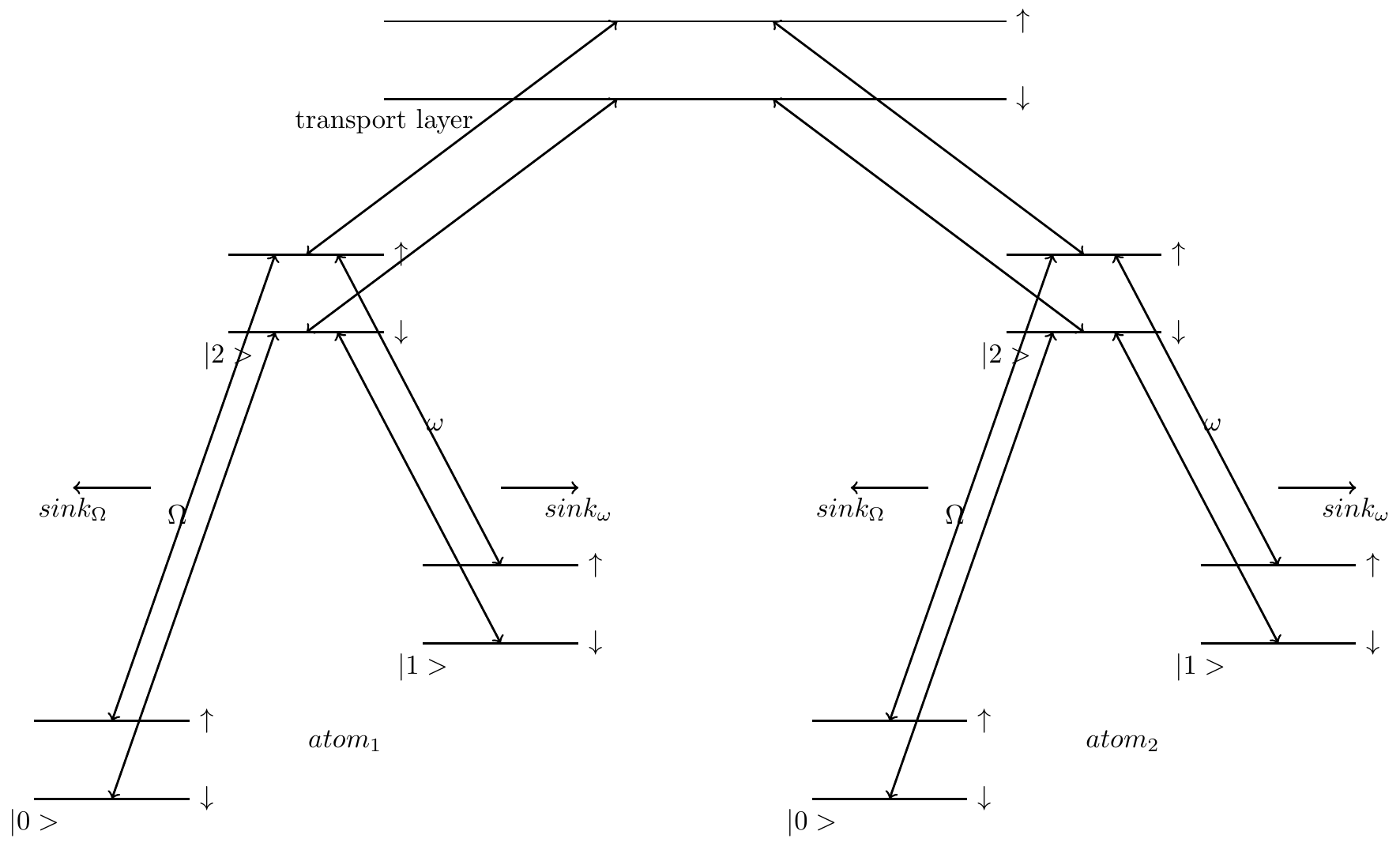}
            \caption{Two atoms with $\lambda$-spectrum}
            \label{lambda-specturm}
        \end{center}
    \end{figure}

    In other modification we took more complex model which included two atoms with $\lambda$-spectrum. System has two electrons with spins: up and down and a transport layer between the atoms. The function of transport layer is to describe the tunnel effect. This means that electrons can be transferred from one atom to another using a transport layer. The system is shown in the Figure \ref{lambda-specturm}.

    The basis of the model can be written as follows: $|o\rangle^{\uparrow}_a\otimes|o\rangle^{\downarrow}_a$ ,where $\uparrow$ - spin up, $\downarrow$ - spin down, $o=0,1,2$ - index of orbit, $a=1,2$ - index of atom.

\begin{figure}[h!]
        \begin{center}
                \includegraphics[width=1.0\textwidth]{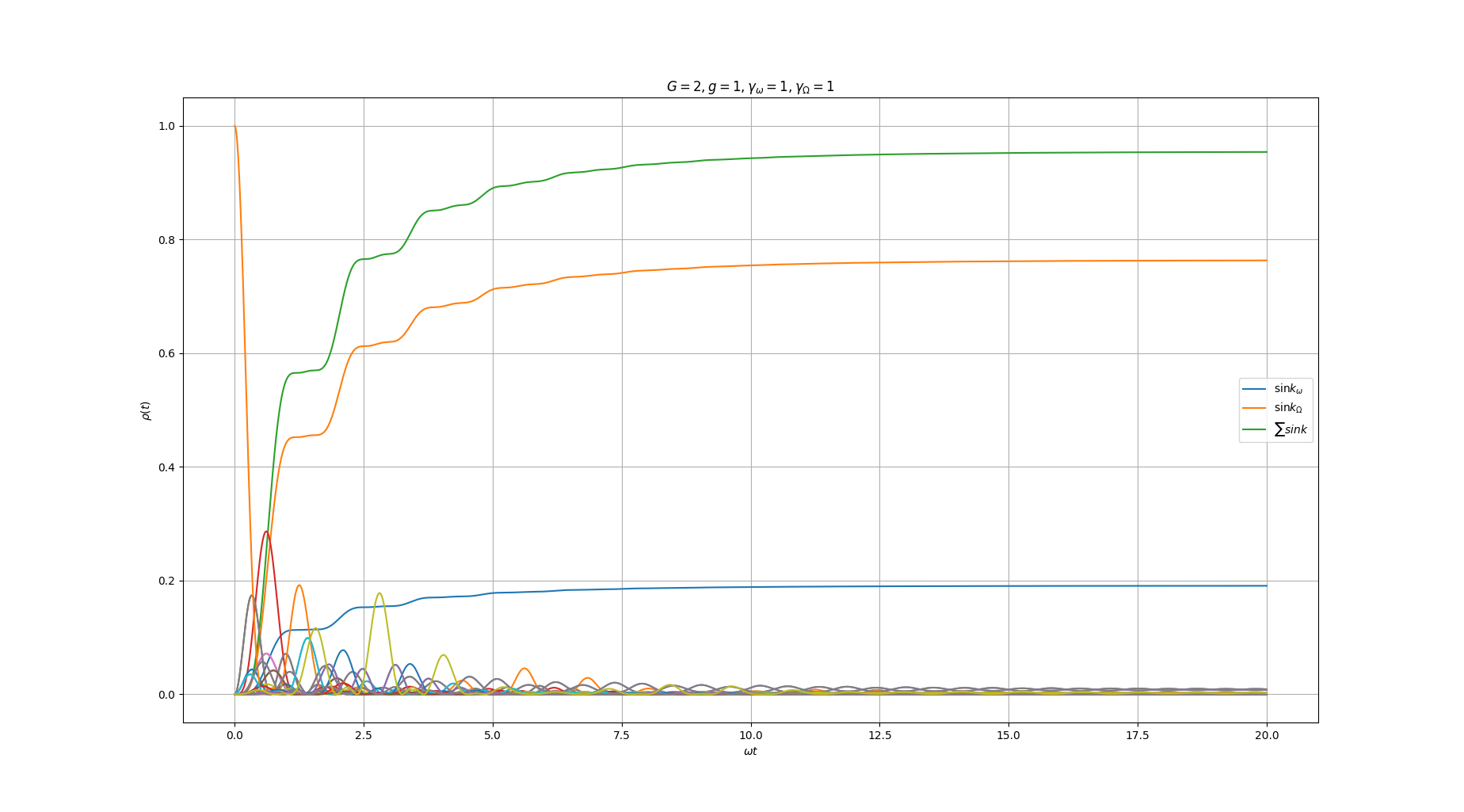}
                \caption{Dynamics of atoms. Sum of probability of $sink_{\omega}$ and $sink_{\Omega}$ is unequal to 1. It means some dark states exist.}
                \label{lambda-specturm-results}
        \end{center}
    \end{figure}

    We spend an experiment with initial state: $|2\rangle^{\uparrow}_1|2\rangle^{\downarrow}_1$ and such parameters: $g_{\Omega}=2$, $g_{\omega}=1$, $\gamma_\omega = 1$, $\gamma_\Omega = 1$. Results of this experiment is shown in the fig. \ref{lambda-specturm-results}.

	At the end of experiment we got the finite dark states:
$$
	    \frac{1}{\sqrt{2}}(|0\rangle^{\uparrow}_1|2\rangle^{\downarrow}_2-|2\rangle^{\uparrow}_1|0\rangle^{\downarrow}_2), \frac{1}{\sqrt{2}}(|1\rangle^{\uparrow}_1|2\rangle^{\downarrow}_2-|2\rangle^{\uparrow}_1|1\rangle^{\downarrow}_2), \frac{1}{\sqrt{2}}(|2\rangle^{\uparrow}_1|2\rangle^{\downarrow}_1-|2\rangle^{\uparrow}_2|2\rangle^{\downarrow}_2)
$$

\newpage

\section{Conclusion}

We modified the Tavis-Cummings-Hubbard scheme for artificial chemistry computions to include the structure of the atom, the electronic transitions, and the motion of the atoms themselves. We have shown that within the framework of such a model, a plausible representation of the association - dissociation reactions of two artificial two-level atoms into an artificial molecule is possible. Both processes are represented as solutions to the same basic quantum equation, but with different initial conditions.

We have shown the possibility of optical interpretation of chemical dynamics for the simplest case of an electron transition between atoms. In this interpretation, the dynamics is represented only in the form of an exchange of photons between multi-level atoms, so that the movement of massive particles is distorted. This makes optical interpretation more convenient by physically implementing chemistry models.

In our modified model, we investigated the electronic transitions between three-level atoms with a spectrum of lambda type, and established the possible dark states of ensembles of a pair of such atoms.

These results indicate the adequacy of our modification of the Tavis-Cummings-Hubbard model and the possibility of its scaling to large ensembles of atoms, including multi-level ones.

\section{Acknowledgements}

The paper was published with the financial support of the Ministry of Education and Science of the Russian Federation as part of the program of the Moscow Center for Fundamental and Applied Mathematics under the agreement №075-15-2019-1621.

\end{document}